\documentclass{article}
\usepackage[utf8]{inputenc}
\usepackage{amsfonts}
\usepackage{amsmath}
\usepackage{graphicx}
\usepackage{geometry}
\usepackage{hyperref}
\usepackage[round]{natbib}
\usepackage{authblk}
\usepackage[shortcuts]{extdash}
\graphicspath{{../figures/},{figures-maosoam/},{./}}

\renewcommand{\cite}[1]{\citep{#1}}

\title{Routes to long-term atmospheric predictability in reduced-order coupled ocean-atmosphere systems -- Impact of the ocean basin boundary conditions}

\author{St\'{e}phane Vannitsem\footnote{svn@meteo.be}}
\author{Roman Sol\'{e}-Pomies}
\author{Lesley De~Cruz}

\affil{Royal Meteorological Institute of Belgium, Avenue Circulaire 3, 1180 Brussels, Belgium}

\date{}

\begin{document}

\maketitle

\begin{abstract}
The predictability of the atmosphere at short and long time scales, associated with the coupling to the ocean, is explored in a new version of the Modular Arbitrary-Order Ocean-Atmosphere Model (MAOOAM), based on a 2-layer quasi-geostrophic atmosphere and a 1-layer reduced-gravity quasi-geostrophic ocean. This version features a new ocean basin geometry with periodic boundary conditions in the zonal direction. The analysis presented in this paper considers a low-order version of the model with 40 dynamical variables.

First the increase of surface friction (and the associated heat flux)  with the ocean can either induce chaos when the aspect ratio between the meridional and zonal directions of the domain of integration is small, or suppress chaos when it is large. This reflects the potentially counter-intuitive role that the ocean can play in the coupled dynamics.

Second, and perhaps more importantly, the emergence of long-term predictability within the atmosphere for specific values of the friction coefficient occurs through intermittent excursions in the vicinity of a (long-period) unstable periodic solution. Once close to this solution the system is predictable for long times, i.e. a few years. The intermittent transition close to this orbit is, however, erratic and probably hard to predict. 

This new route to long-term predictability contrasts with the one found in the closed ocean-basin low-order version of MAOOAM, in which the chaotic solution is permanently wandering in the vicinity of an unstable periodic orbit for specific values of the friction coefficient. The model solution is thus at any time influenced by the unstable periodic orbit and inherits from its long-term predictability.
\end{abstract}

\section{Introduction}

Nowadays important efforts are devoted to the development of forecasting systems for long lead times from seasons to decades. 
Just like numerical weather prediction (NWP) systems, these forecasts are subject to the property of sensitivity to initial conditions 
by which a small initial error will rapidly amplify until it reaches a level at which no useful information can be extracted anymore. 
This level is usually fixed at the point where the mean square error score reaches twice the variance of the climatology of the observable  \citep{Lorenz1982,Dalcher1987,Chen1989,Boer1994,Savijarvi1995}.
This specific level marks the time at which, on average, no correlation exists between the forecast and the observations anymore.
So the way the mean square error reaches this level as a function of time is a key feature revealing the potential of extended-range
forecasts.  

This feature was already recognized by \citet{Lorenz1982} who estimated the limit of predictability of weather systems at mid-latitudes to 10-15 days 
using a state-of-the-art weather forecasting system (at that time). Similar estimates have been obtained later in more sophisticated NWP models
which only describe the evolution of the atmosphere \citep{Dalcher1987, Chen1989, Kalnay2003}. Nowadays, the tendency is to develop Earth System Dynamic (ESD) models encompassing different components
of the climate system. The hope is that the mutual interaction between
these different components will allow to extend the predictability limit of the atmosphere beyond the 10-15 days barrier.    

Interactions between the different components of the climate system are essential to its behaviour on multiple time scales. A particular example is the
dynamics of the El Ni\~no--Southern Oscillation (ENSO) phenomenon for which the ocean interacts with the atmosphere in the tropical Pacific, leading to the development of
a coupled dynamics with a low-frequency variability developing in the atmosphere on time scales of a few years, e.g. \cite{Philander1990, Stuecker2013}. At mid-latitudes, such ocean-atmosphere 
interactions are less pronounced
and the origin of the low-frequency variability is still a matter of debate \citep{Robertson2000, Marshall2001}. Some coupled modes have however been found in coupled ocean-atmosphere models or in observational data
\citep{Czaja2002, VanderAvoird2002, Kravtsov2007, Feliks2007, Minobe2008, Brachet2012, Lheveder2015, Vannitsem2017a}. This suggests that low-frequency variability developing in the 
atmosphere can be associated with the coupling to the ocean, and could be a source of long-term predictability.  

Recently a simple  reduced-order multi-scale system composed of a reduced-gravity quasi\-/geostrophic ocean layer and a two-layer baroclinic atmosphere has been built to sketch the dynamics
of the coupled ocean-atmosphere system over mid-latitude ocean basins. The fields described by these equations were developed in Fourier series and severely truncated to a low order, yielding
a system of 36 ordinary differential equations describing the dominant modes of the dynamics \cite{Vannitsem2015a}.  This system displays
a low-frequency variability (LFV) in the atmosphere provided the friction coefficient between the ocean and the atmosphere is sufficiently high \cite{Vannitsem2015a}. The emergence
of this low-frequency variability is at the origin of the long-term predictability of some observable within the atmosphere as illustrated in \cite{Vannitsem2017b}. It is however
important to notice here that in the context of this coupled model, the LFV does not develop when the friction coupling (which also controls the heat fluxes between the two
components) is small. This means that coupling two components, slow and fast, does not imply that the fast system will inherit the LFV associated with the presence of the slow component.
In the model version of \citet{Vannitsem2015a}, called VDDG in the following, the development of the LFV and the long-term predictability in the atmosphere is the outcome of a bifurcation leading to a drastic
qualitative change of dynamics. The solution of the system experiences a catastrophic change when the friction coefficient is increased, leading to an attractor 
developing close to an unstable periodic orbit.  We will refer to this transition as a {\it chaos-to-chaos transition}. This new solution wanders chaotically
around the unstable periodic orbit, a type of dynamics that has already been isolated in much simpler systems by e.g. \citet{Tel2006, Wernecke2017}.  
This chaos-to-chaos transition has been found in autonomous and non-autonomous versions of the model \cite{Vannitsem2015b}, and 
in all the experiments that were done with this model configuration \citep{Vannitsem2015a, Vannitsem2015b, Vannitsem2017b, Demaeyer2018}.         

One can now wonder whether the presence of this transition is sensitive to changes in the model, such as an increase in the model resolution or a change in the geometry of the system. This has motivated the development of a flexible, generalised version of the model called the Modular Arbitrary-Order Ocean-Atmosphere Model (MAOOAM) \citep{DeCruz2016}. Using MAOOAM, some preliminary work has been done in answering the first question, by looking at the variability of the solutions when increasing the number of modes.  It was shown that
the LFV can weaken at intermediate resolutions, but recovers as the number of modes is further increased.

This work aims to address the second question by modifying the basin boundary conditions of the ocean. Instead of using a fully closed ocean basin, periodic boundary conditions are imposed in the zonal direction of the ocean component. In this situation, both the flow in the ocean and in the atmosphere develop in a channel.   

This new model configuration can be viewed as a very crude representation of the Southern Ocean encircling Antarctica. However, care should be taken when using the results presented in the following sections to interpret the dynamical behaviour in the Southern Ocean. Indeed, the model does not feature a meridional gradient of density in the ocean component, which is composed of a single homogeneous layer. This gradient is known to be essential in the development of the Antarctic Circumpolar Current (ACC), \cite{Vallis2006}. It is nonetheless interesting to investigate the robustness of the findings obtained in the context of a closed ocean basin in this new model version. This analysis can provide new ideas in the analysis of the dynamics of a more realistic
Southern Ocean configuration. 

In Section 2, the new model version is described. The Lyapunov properties characterizing the short-term evolution and the long-term saturation of the error within this model
are then discussed in Section 3. The key conclusions are provided in Section 4. 

\section{The new version of the model}

\subsection{Dynamical equations}
\label{sec:dyn}

The dynamical equations of the model have already been described in \citet{Vannitsem2015b, DeCruz2016}. These are briefly repeated here for completeness. 

\subsubsection{Equations of motion for the atmosphere and the ocean}
\label{ssec:atmos}

The atmospheric model is based on the vorticity equations of a two-layer, quasi-geostrophic
flow defined on a $\beta$-plane. The equations in pressure coordinates are
\begin{eqnarray}
\frac{\partial}{\partial t} \left( \nabla^2 \psi^1_a \right) + J(\psi^1_a, \nabla^2 \psi^1_a) + \beta \frac{\partial \psi^1_a}{\partial x}
& = & -k'_d \nabla^2 (\psi^1_a-\psi^3_a) + \frac{f_0}{\Delta p} \omega, \nonumber \\
\frac{\partial}{\partial t} \left( \nabla^2 \psi^3_a \right) + J(\psi^3_a, \nabla^2 \psi^3_a) + \beta \frac{\partial \psi^3_a}{\partial x}
& = & +k'_d \nabla^2 (\psi^1_a-\psi^3_a) - \frac{f_0}{\Delta p}  \omega \nonumber \\  
& & - k_d \nabla^2 (\psi^3_a-\psi_o); 
\label{eq:atmos}
\end{eqnarray}
here $\psi^1_a$ and $\psi^3_a$
are the streamfunction fields at 250 and 750 hPa, respectively, $\omega = dp/dt$ is the vertical velocity,
$f_0$ is the Coriolis parameter at latitude $\phi_0$, and $\beta = df/dy$ the meridional gradient of $f$ at latitude $\phi_0$.

The coefficients $k_d$ and $k'_d$
multiply the surface friction term and the internal friction between the layers, respectively, while $\Delta p = 500$ hPa is the pressure
difference between the two atmospheric layers. An additional term has been introduced in this system in order to account for the presence of a
surface boundary velocity of the oceanic flow defined by $\psi_o$ whose evolution is based on 
the reduced-gravity, quasi-geostrophic shallow-water model on a $\beta$-plane :
\begin{equation}\label{eq:QGSW}
\frac{\partial}{\partial t} \left( \nabla^2 \psi_o - \frac{\psi_o}{L_R^2} \right) + J(\psi_o, \nabla^2 \psi_o) + \beta \frac{\partial \psi_o}{\partial x}
= -r \nabla^2 \psi_o + \frac{{\mathrm{curl}}_z \tau}{\rho h}.
\end{equation}
describing the dynamics within the model ocean's upper, active layer, where $\rho$ is the density of water of the upper layer, $h$ the depth of
this layer, $L_R$ the reduced Rossby deformation radius, $r$ a friction coefficient at the bottom of the
active layer, and ${\mathrm{curl}}_z \tau$ is the vertical component of the curl of the wind stress that will be defined later.

\subsubsection{Ocean temperature equation}
\label{ssec:ocean_temps}

We assume that temperature is a passive scalar transported by the
ocean currents, but the oceanic temperature field
displays strong interactions with the atmospheric temperature through radiative and heat
exchanges.  Under these assumptions, the evolution equation for the ocean temperature is

\begin{equation}\label{eq:heat_oc}
\gamma_o ( \frac{\partial T_o}{\partial t} + J(\psi_o, T_o)) = -\lambda (T_o-T_a) + E_R
\end{equation}
with
\begin{equation}\label{eq:fluxes_oc}
E_R = -\sigma_B T_o^4 + \epsilon_a \sigma_B T_a^4 + R_o.
\end{equation}

In Equations \eqref{eq:heat_oc} and \eqref{eq:fluxes_oc} above, $E_R$ is the net radiative flux at the ocean surface (positive into the ocean), $R_o$ is
the shortwave radiation entering the ocean, $\epsilon_a$ the emissivity of the atmosphere, $\sigma_B$ the Stefan-Boltzman constant, $\gamma_o$
the heat capacity of the ocean, and $\lambda$ is the heat transfer coefficient between the ocean and the atmosphere
that combines both the latent and sensible heat fluxes. It is assumed that this combined heat transfer is
proportional to the temperature difference between the atmosphere and the ocean.

\subsubsection{Atmospheric temperature equation}
\label{ssec:atmos_temps}

The thermodynamic equation for the atmosphere is written as,

\begin{equation}\label{eq:heat_atm}
\gamma_a ( \frac{\partial T_a}{\partial t} + J(\psi_a, T_a) -\sigma \omega \frac{p}{R}) = -\lambda (T_a-T_o) + E_{a,R}
\end{equation}
with
\begin{equation}\label{eq:fluxes_atm}
E_{a,R} = \epsilon_a \sigma_B T_o^4 - 2 \epsilon_a \sigma_B T_a^4 + R_a.
\end{equation}

In these two equations,
$R$ is the gas constant, and
$$\sigma =  - \frac{R}{p} \Big(\frac{\partial T_a}{\partial p}- \frac{1}{\rho_a c_p}\Big) $$
is the static stability, with $p$ the
pressure, $\rho_a$ the air density, and $c_p$ the specific heat at constant pressure;  here $\sigma$ is
taken to be constant. $R_a(t)$ represents the portion of the short-wave radiative input from the sun directly captured by the atmosphere. 

Note also that, thanks to the hydrostatic relation in pressure coordinates and to the
ideal gas relation $p=\rho_a R T_a$, the atmospheric temperature $T_a$ can be expressed as
$T_a = - (p/R) f_0 (\partial \psi_a/\partial p)$.  This expression for $T_a$ can then be used to combine Equations \eqref{eq:heat_atm} and \eqref{eq:atmos},
as done when deducing the quasi-geostrophic potential vorticity equation \citep[e.g.][]{Vallis2006}.

\subsection{Domain of integration and parameter values}
\label{sec:maosoamdescription}

In the original model version of MAOOAM, the domain of integration is rectangular, with a closed ocean basin and 
a channel flow for the atmosphere \citep{Vannitsem2015a, DeCruz2016}. Free-slip boundary conditions were chosen along the meridional and
longitudinal boundaries of the ocean basin, while these are free-slip in the meridional direction and periodic in the zonal
direction for the atmosphere. In the new version of the model, the main modification    
is that we now impose periodic boundary conditions in the zonal direction in the ocean. 
This model now represents a channel flow for both the ocean and the atmosphere.

These boundary conditions are satisfied by using the basis functions that were previously used for the atmosphere to expand both the atmosphere and the ocean fields. In other words, with the proper normalization, the basis functions for all fields must be of the following form, following the nomenclature of
\citet{Cehelsky1987}:
\begin{align}
&F^A_{P} (x', y')   =  \sqrt{2}\, \cos(P y')\label{eqn:FA}\\
&F^K_{M,P} (x', y') =  2\cos(M nx')\, \sin(P y')\label{eqn:FK}\\
&F^L_{H,P} (x', y') = 2\sin(H nx')\, \sin(P y'),\label{eqn:FL}
\end{align}
with $(P, M, H) \in \mathbb{N}^3$. The set of modes that will be used in the present study are such that $P=1, 2$, $M=1, 2$ and $H=1, 2$, implying that
10 modes will be used for the four dynamical fields, $\psi_a=\sum_{i=1}^{10} \psi_{a,i} F_i$, $\theta_a=\sum_{i=1}^{10} \theta_{a,i} F_i$, $\psi_o=\sum_{i=1}^{10} \psi_{o,i} F_i$ and 
$T_o=\sum_{i=1}^{10} T_{o,i} F_i$, where $F_i$ are simplified notations for the set of modes used. Note that the first mode is $ F_1=\sqrt{2}\, \cos(y')$ whose coefficients are denoted as  $\psi_{a,1}$, $\psi_{o,1}$ $T_{o,1}$. These variables will be analyzed in detail in the next section.  

Let us now estimate the parameters to some realistic values \citep{Vannitsem2015b}. 
Assuming that the wind stress follows the linear relation, $(\tau_x, \tau_y)=C (u-U,v-V)$ -- where  $(u = -\partial \psi_a/\partial y, v = \partial \psi_a/\partial x)$ are the horizontal 
components of the geostrophic wind, respectively, and $(U, V)$ the corresponding components of the geostrophic currents in the ocean -- one gets,
\begin{equation}
\frac{\mathrm{curl}_z \tau}{\rho h} = \frac{C}{\rho h} \nabla^2 (\psi^3_a-\psi_o). 
\label{eq:stress}
\end{equation}
where $C$=$\rho_a C_D |\vec{v}-\vec{V}|$ with $C_D$ the drag coefficient and $|\vec{v}|$, the norm of the velocity.
The coefficient $d = C/(\rho_o h)$ characterizes the strength of the mechanical coupling between the ocean and the atmosphere.

Similarly, one can use the Ekman layer theory to relate the coefficient $k_d$ in Equation \ref{eq:atmos} to the friction coefficient $C$ 
in pressure coordinates,
\begin{equation}
k_d=\frac{g C}{\Delta p} \, [s^{-1}]
\end{equation}
and one assumes that $k'_d$=$k_d$ as in \cite{Charney1980}.

The parameter $\lambda$ in Equation \ref{eq:heat_oc} can also be related to
the surface friction coefficient as discussed in \citet{Houghton1986},
\begin{equation}
\lambda= 1004 \, C  \, [W m^{-2} K^{-1}]
\label{lambda}
\end{equation}

The radiative input is decomposed in two different terms, $R_o=R_{o,0}+\delta R_{o}$, the first one being a constant value in space
and the second one a meridionally dependent term $\delta R_{o} = C_o F_1$.  

The parameter values used in the present analysis are given in Table \ref{tab:par}. 

\begin{table}
\caption{List of parameters of the model\label{tab:par}}
\begin{tabular}{lr|lr}
\hline

    Parameter (unit) & Value & Parameter (unit) & Value \\
\hline
    $L_y = \pi L$  (km)         & $5.0 \times 10^3$        & $\gamma_\text{o}$ (J\,m$^{-2}$\,K$^{-1}$) & $4 \times 10^6 \, h$  \\
    $f_0$ (s$^{-1}$)            & $1.195 \times 10^{-4}$   & $C_\text{o}$ (W\,m$^{-2}$)             & Variable \\
    $n = 2 L_y / L_x$           & Variable                 & $T_\text{o}^-2$ (K)                     & $285$ \\
    $R_\text{E}$ (km)           & $6370$                   & $\gamma_\text{a}$ (J\,m$^{-2}$\,K$^{-1}$) & $1.0 \times 10^7$ \\
    $\phi_0$                    & $-0.3056 \pi$            & $C_\text{a}$ (W\,m$^{-2}$)             & $C_\text{o}/3$ \\
    $g^\prime$                  & $3.1 \times 10^{-2}$     & $\epsilon_\text{a}$                    & $0.76$ \\
    $r$ (s$^{-1}$)              & $1.0 \times 10^{-8}$     & $\beta$ (m$^{-1}$\,s$^{-1}$)           & $1.62 \times 10^{-11}$ \\
    $h$ (m)                     & Variable                    & $T_\text{a}^0$ (K)                     & $270$ \\
    $d$ (s$^{-1}$)              & $C/(\rho_o h)$     & $\lambda$ (W\,m$^{-2}$\,K$^{-1}$)      &   $1004 \, C$ \\
    $k_d$ (s$^{-1}$)            & $\frac{g C}{\Delta p}$    & $R$ (J\,kg$^{-1}$\,K$^{-1}$)           & $287$ \\
    $k_d^\prime $ (s$^{-1}$)    & $\frac{g C}{\Delta p}$     & $\sigma$  (J kg$^{-1}$ Pa$^{-2}$)                & $2.16 \times 10^{-6}$ \\   
    $C$ (kg m$^{-2}$ s$^{-1}$)       & Variable & & \\ 
 \hline
  \end{tabular}
  \end{table}
Four important parameters will be modified in the current investigation, $C$, $n$, $h$ and $C_o$.

\subsection{Typical solutions of the model}
\label{solution}

\begin{figure}
\centering
\includegraphics[width=0.90\textwidth]{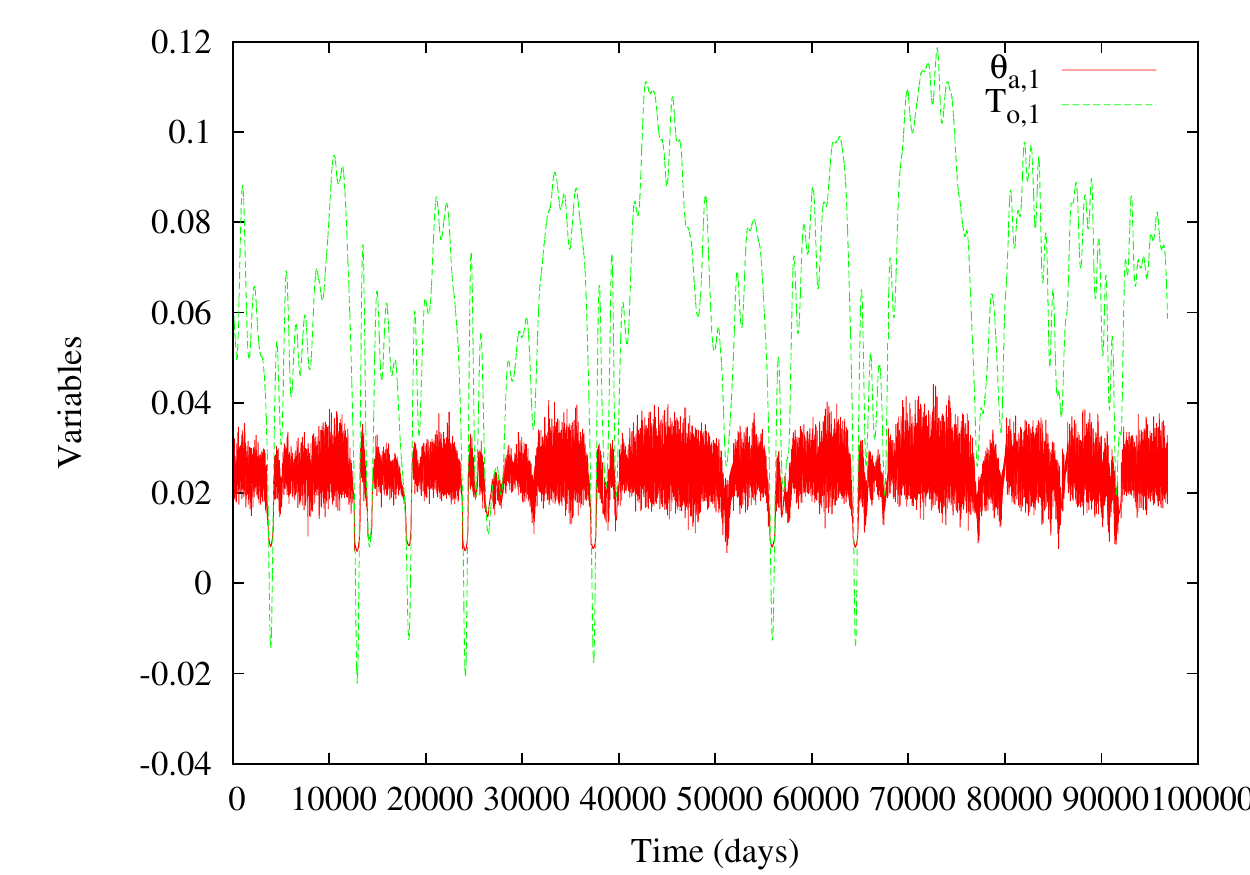}
\caption{Time series of $\theta_{a,1}$ and $T_{o,1}$ for $C=0.016$ kg m$^{-2}$ s$^{-1}$, $n=1.7$, $C_o=350$ W m$^{-2}$ and $h=1000$ m. Nondimensional units are used for the variables.}
\label{Figseries}
\end{figure}

Let us first briefly qualitatively analyze the solutions generated by the coupled model. Time series of  
$\theta_{a,1}$ and $T_{o,1}$ for $C=0.016$ kg m$^{-2}$ s$^{-1}$, $n=1.7$, $C_o=350$ W m$^{-2}$ and $h=1000$ m, are displayed
in Fig. \ref{Figseries} for about 100,000 days. A first remarkable result is the difference of typical time scale of variability. 
For the first mode of temperature in the ocean, a low-frequency variability is visible on time scales of the order of 10,000 days; 
while for the first temperature atmospheric mode, $\theta_{a,1}$, a high frequency variability on a time scale of days is present together with a low-frequency variability coherent with the ocean temperature evolution. This evolution suggests that the dynamics
within the atmosphere is influenced by the ocean, and should therefore lead to modifications of the predictability properties of the
atmosphere. 

In order to further clarify the behaviour of the solutions of the system in this range of values, several videos have been prepared and 
are 
provided as supplementary material (videos S1-S7). These videos display the solutions generated by the model for $h=1000$ m and two values of
$n=1.5$ and $1.7$, and several values of $C$. Six different panels are displayed in each video, the geopotential height difference
between two locations in the spatial domain ($x'=0$, $y'=3\pi/4$ and $y'=\pi/4$), a three-dimensional projection of the attractor of the system for the
variables ($\psi_{a,1}$, $\psi_{o,1}$ $T_{o,1}$), and four panels representing the solution in the spatial domain. The spatial view
is centered at the South pole.  

Let us focus for the moment on the solutions
obtained with $n=1.7$. For $C=0.01$ kg m$^{-2}$ s$^{-1}$ (video S1), the temporal evolution of the geopotential height displays an erratic behaviour 
without any visible low-frequency variability. This erratic behaviour is also visible in the three-dimensional plot. The solutions
in space display a rapid erratic dynamics for the atmospheric temperature (proportional to the baroclinic streamfunction) and 
barotropic atmospheric streamfunction, while a much slower evolution is seen for the ocean fields. Interestingly a slow 
counterclockwise motion is found for the ocean streamfunction, with a westward Rossby wave-type of motion. 
Overall the dynamics of the atmosphere seems to be well independent of the dynamics of the ocean.             

Let us now increase the friction coefficient to $C=0.016$ kg m$^{-2}$ s$^{-1}$ (video S2). In this case, the dynamics displays an intermittent behaviour
with periods of strong variability and periods during which the dynamics seems frozen. It is particularly spectacular for the
atmospheric streamfunction which seems to be frozen for a while from time to time. In the three-dimensional projection of the
attractor this dynamics is marked by a trajectory (the red dot) exploring high values of $\psi_{o,1}$ and small values of $\theta_{o,1}$. 
The trajectory seems to converge to a domain of the state space associated with a purely periodic motion 
which is visible in the counterclockwise rotation of the spatial ocean streamfunction field. For even larger values of $C$, a similar
picture is found as illustrated in the videos S3 and S4 for $C=0.02$ and $C=0.027$ kg m$^{-2}$ s$^{-1}$, respectively. In the latter case, the almost periodic motion is 
clearly visible alternating with more erratic behaviour, in particular in the geopotential height. 
This almost periodic motion has quite a long period of about 10 years for $C=0.027$ kg m$^{-2}$ s$^{-1}$.

These videos illustrate the important role of the ocean in the development of low-frequency variability within the atmosphere. A first
interesting aspect is the fact that the atmosphere does not "feel" the ocean if the friction is not large enough (also related to the
heat fluxes between the two components through Equation (\ref{lambda})). When a sufficiently high value of the friction coefficient $C$ is reached, the solution experiences intermittent excursions toward a situation for which the dynamics of the atmosphere and the ocean
are coherent with each other. These different types of behaviour for small and large values of $C$ should play a role in the predictability
of the atmosphere. This question will now be addressed by investigating two quantities, the Lyapunov exponents and the error dynamics.

\section{Predictability properties of the model}

As illustrated in section \ref{solution}, the solutions of the model can display erratic types of behaviour reminiscent of the dynamics of chaotic deterministic dynamical systems. The predictability of such systems is limited in time as a small error in the initial conditions will
grow in time until it reaches a level at which the forecast cannot be distinguished from a random draw of a solution on the attractor 
of the system (a draw from the long-term climatology of the system). This property of sensitivity to initial conditions can be evaluated by computing the Lyapunov exponents and/or by computing the evolution of the error during the entire period of the forecasts. The first approach assumes
infinitesimal initial errors and is commonly used to characterize the nature of the dynamics such as stationary, periodic or chaotic
solutions. The second approach does not assume that the initial errors are infinitesimal, which is of course more appropriate when
dealing with realistic forecasting problems. Both will be used here, first to characterize the nature of the solutions, and second
to evaluate the long-term predictability of the flow.

\subsection{Lyapunov instability of the model}

Let us first briefly define these quantities and then look at the exponents obtained in the present model. More details can be found in e.g. \citet{Kuptsov2012}. 

\subsubsection{Lyapunov exponents}

The evolution laws of a dynamical system like the ones presented in Section \ref{sec:dyn} can be written in the compact form
\begin{equation}
\frac{d\vec{x}}{dt} = {\vec f}(\vec x, \lambda)
\label{equat}
\end{equation}
where $\vec x$ is a vector containing the set of relevant variables $\vec x$ = $( x_1, ..., x_n)$. This system of equations
is then integrated in time starting from an initial state, $\vec x (t_0)=\vec x_0$. As the real state $\vec x_0$ is never known with
infinite precision in practice, a small error, $\delta \vec x_0$, will affect the future forecast. This perturbed initial state 
generates a new trajectory in phase space. The time-dependent error vector, that is the displacement vector between the reference trajectory and 
the perturbed one at a given time, is denoted $\delta \vec x (t)$. Provided the initial error is small, its evolution is described
by a linearized system of equations,
\begin{equation}
\frac{d\delta \vec x}{dt} =  \frac{\partial \vec f}{\partial \vec x}_{\vert \vec x(t)}
\delta \vec x
\label {linear}
\end{equation}
whose formal solution is,
\begin{equation}
\delta \vec x (t) = {\bf M}(t,\vec{x}(t_0)) \delta \vec x (t_0)
\end{equation}
where the matrix $\bf{M}$ is referred as the resolvent matrix. In the ergodic theory of deterministic dynamical systems, the double
limit of infinitely small initial errors and infinitely long times, is usually taken, e.g. \citet{Eckmann1985}. In
this limit the divergence of initially close states is determined by the logarithm of the eigenvalues of the
matrix $(\bf{M}^T \bf{M})^{2(t-t_0)}$ that are referred to as the Lyapunov exponents.
The full set of Lyapunov exponents of a system is called the Lyapunov spectrum, which is usually represented in
decreasing order. Positive Lyapunov exponents indicate the presence of a chaotic dynamics, hence displaying sensitivity to initial conditions.

Several numerical techniques have been developed to evaluate these Lyapunov exponents \cite{Parker1989, Kuptsov2012}.
One of the most popular methods consists in following the evolution of a set of
orthonormal vectors chosen initially at random in the tangent space of the trajectory
$\vec x(t)$. This basis is regularly orthonormalized using the standard Gram-Schmidt method to avoid the
alignment of all the vectors along the unstable direction associated to the largest Lyapunov
exponent, and the amplification along these vectors can then be computed. The logarithm of these amplifications are
then computed and rescaled by the time $t-t_0$, which provides the set of Lyapunov exponents.
This is the approach adopted here.

\subsubsection{Results for MAOOAM with a channel ocean}

\begin{figure}
\centering
\includegraphics[width=1\textwidth]{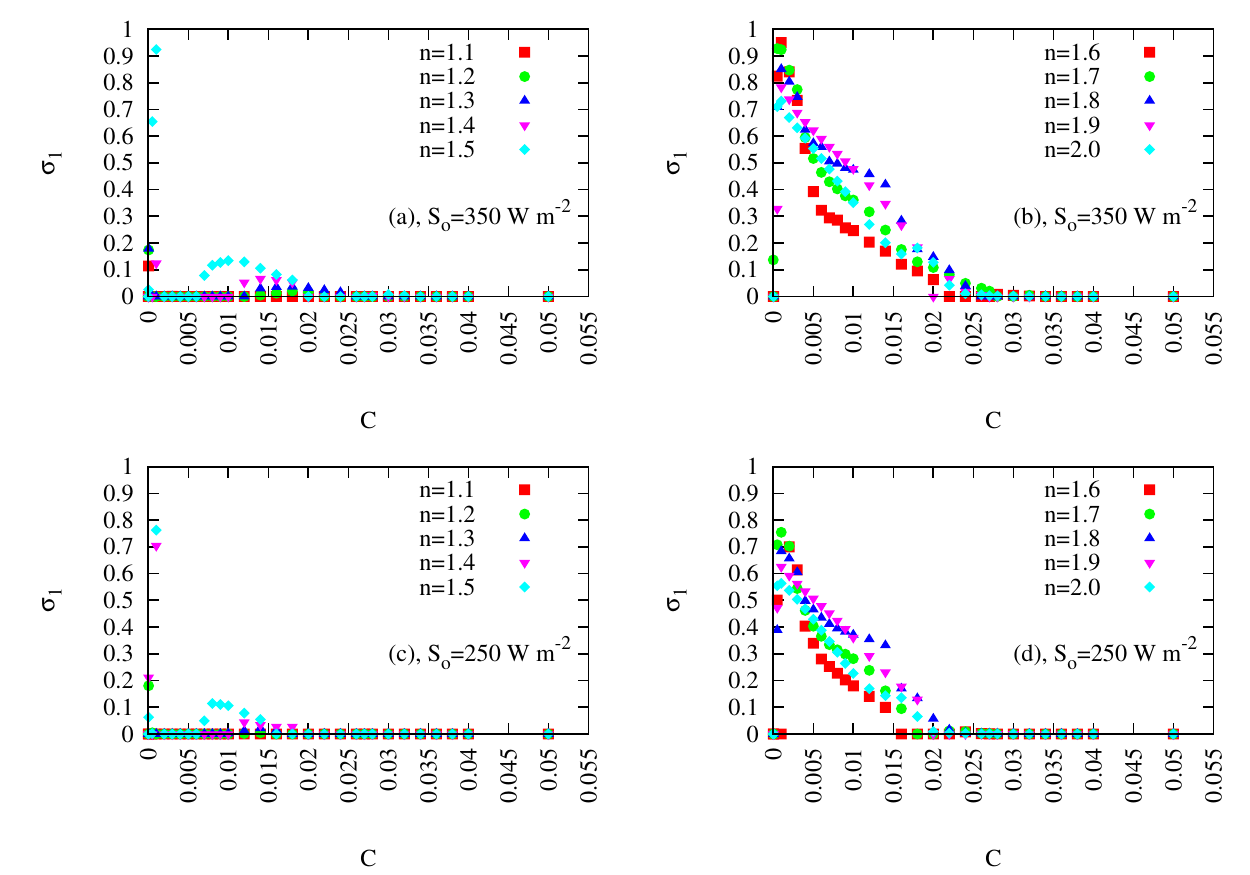}
\caption{First Lyapunov exponent as a function of the friction coefficient $C$ [kg m$^{-2}$ s$^{-1}$] for (a) different values of the aspect ratio $n$ and for $C_o=350$ W m$^{-2}$ and $h=1000$ m; (b)
as in (a) but with larger values of $n$; (c) as in (a) but for $C_o=250$ W m$^{-2}$; and (d) as in (c) but for $C_o=250$ W m$^{-2}$.}
\label{Figlyap1}
\end{figure}

Figure \ref{Figlyap1} displays the dependence of the first Lyapunov exponent as a function of the friction coefficient $C$ for $C_o=350$ and $C_o=250$
W m$^{-2}$ with $h=1000$ m. Different values of the aspect ratio, $n$, are considered in the different panels. Let us start with small values of the
aspect ratio in panel (a) for $C_o=350$ W m$^{-2}$. For very small values of $n$, the solutions found are periodic, leading to a first Lyapunov exponent
equal to 0. When $n$ is increased, the solutions become chaotic for an intermediate set of values of the friction coefficient, $C$. This indicates that the coupling between
the ocean and the atmosphere through surface friction can induce a chaotic dynamics, a regime referred to as {\it ocean-induced chaotic dynamics}. For even larger
values of $n$ (panel b), the picture is different, with a decrease of the value of the dominant Lyapunov exponents as a function of the friction coefficient, suggesting
a stabilization of the flow through friction. This feature is similar to the dependence found for the original VDDG model with $n=1.5$ \citep{Vannitsem2017b}. 
This regime can be referred to as a {\it ocean-tempered chaotic dynamics}. A similar picture is found when the solar input is reduced to $C_o=250$ W m$^{-2}$ as illustrated in panels (c) and (d).    

\begin{figure}
\centering
\includegraphics[width=1\textwidth]{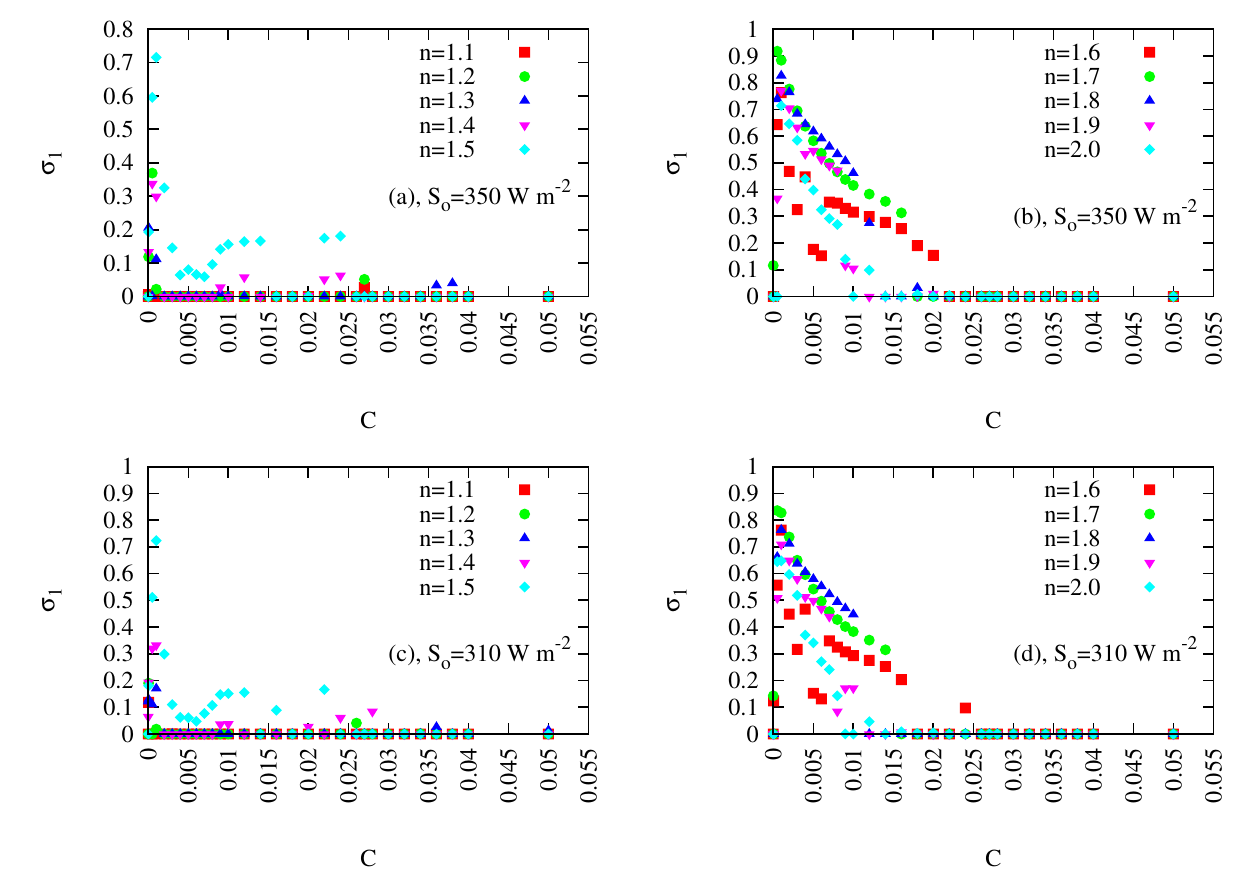}
\caption{First Lyapunov exponent as a function of the friction coefficient $C$ [kg m$^{-2}$ s$^{-1}$] for (a) different values of the aspect ratio $n$ and for $C_o=350$ W m$^{-2}$ and $h=100$ m; (b)
as in (a) but with larger values of $n$; (c) as in (a) but for $C_o=310$ W m$^{-2}$; and (d) as in (c) but for $C_o=310$ W m$^{-2}$.}
\label{Figlyap2}
\end{figure}

When the depth of the ocean layer is reduced to $h=100$ m, shown in Figure \ref{Figlyap2}, a qualitatively similar picture is found, but with a more complicated dependency structure of
the friction coefficient $C$. For instance with $n=1.5$, alternating windows of chaotic and periodic solutions are present, which was not visible for $h=1000$ m at 
panels (a) and (c) in Figure \ref{Figlyap1}.  

In summary, the contrasting behaviour as a function of $n$ reveals a complicated dependence of the dynamics of the coupled system with respect to the size of the domain (measured by $n$) that can lead
to ocean-induced chaos or on the contrary to ocean-tempered chaos when friction is increased. This remarkable result reveals the important potential role played by the ocean, and the  nontrivial dependence of this role  on the resolved dynamics in the ocean. Indeed, increasing the aspect ratio can also be interpreted as filtering out the large-scale zonal waves in favour of an enhanced description of smaller-scale phenomena, shown to be important for the ocean \cite{DeCruz2018}.
It is however not clear yet whether this result is robust when the number of modes is explicitly increased. This question will be addressed in the future.  

\subsection{Error dynamics}

\subsubsection{Definition of the error}

In the previous section, the emphasis was placed on the analysis of the Lyapunov exponents characterizing
the short term evolution of small initial errors. This analysis, although very useful when looking at the emergence
of chaos, should however be complemented by the analysis of the error dynamics when interested in the long-term
predictability of the system. 

Let us consider a solution of the system in phase space, $\vec{x}(t_0)=\vec{x}_0$, at time $t_0$. Observations of this system are affected by finite-amplitude
initial errors that can be
for simplicity be considered as a Gaussian white noise, $\vec{\epsilon}$, and the observed state is then, $\vec{x}'(t_0)=\vec{x}_0+\vec{\epsilon}$. 
One can now measure the error evolution starting from these two initial conditions as
\begin{equation}
\vec{E}(t)=\vec{x}'(t)-\vec{x}(t)
\end{equation}
where $\vec{x}'(t)$ and $\vec{x}(t)$ are the two trajectories starting from the two initial conditions $\vec{x}'(t_0)$ and $\vec{x}(t_0)$ of the perturbed and
unperturbed trajectories. Since the amplification of this error fluctuates on the inhomogeneous attractor of the system, an average
over the attractor is necessary in order to obtain properties that are independent of the initial state. We do not use the classical norm ($L^2$ norm) 
but rather the logarithmic norm \cite{Nicolis1995}, 
\begin{equation}
\langle \ln E_t^2 \rangle = \int \mathrm{d}\vec{\epsilon}_0 \rho_{\epsilon} (\vec{\epsilon}_0) \int \mathrm{d}\vec{x}_0 \rho_{S} (\vec{x}_0) \ln \left[ (\vec{x}'(t) - \vec{x}(t))  \cdot (\vec{x}'(t)) - \vec{x}(t)) \right]
\label{expnorm0}
\end{equation}
where $\rho_{\epsilon}(\vec{\epsilon}_0) \mathrm{d}\vec{\epsilon}_0$ and $\mathrm{d}\vec{x}_0 \rho_{S} (\vec{x}_0)$, are the probability measure of the initial errors and of the initial 
conditions on the attractor of the system. This specific norm is chosen on the one hand because it is closely related to the definition of the Lyapunov exponents, and
on the other hand because it considerably reduces the fluctuations associated with the finite number of realizations used to compute the mean error. 
The amplitude of the perturbations $\vec{\epsilon}$ is taken sufficiently small in order to get information of the full error growth evolution, even in the short
term regime for which the error evolution follows a linearized system of equations, used to define the Lyapunov exponents. The exponential error dynamics expected
in this linearized evolution regime translates, after a short transient evolution, to a linear error amplification in the logarithmic norm defined above. 
As the error (\ref{expnorm0}) reaches a substantial amplitude, the effect
of the nonlinear terms on the dynamics cannot be neglected anymore and the rate of amplification of the logarithm of the error starts to decrease, and for long lead times, 
saturates due to the finite size of the system's attractor. This evolution is discussed in detail in \cite{Nicolis1995, Vannitsem1994, Vannitsem2016}.      

As we are interested in the long-term predictability of the atmosphere in the coupled system, the focus is placed on how the error defined in (\ref{expnorm0}) 
saturates for a long lead time.  Once this saturation is reached, no predictability is left in the system anymore, see e.g. \cite{Vannitsem2017b}. In other words,  
any random state taken on the attractor of the system will provide a skill score comparable to the one obtained with a set of initialized forecasts for long times (when reaching the saturation level). 

In \cite{Vannitsem2016, Vannitsem2017b} the dynamics of the error is analyzed for the VDDG model version and it is shown that provided the friction is sufficiently high, the
error will continue to grow for very long times (up to 100 years) for certain atmospheric variables. As discussed in the latter references, this remarkable result is 
associated with the development of the low-frequency variability (LFV) into the coupled ocean-atmosphere system through a complex sequence of bifurcations when $C_o$ is increased,
for a sufficiently large value of $C$. Beyond this sequence of bifurcations the attractor of the system develops around an unstable periodic orbit with a very long period.
This orbit constitutes the backbone of the attractor of the system and controls the long term evolution of the error. Let us now investigate how the error behaves in the
new model version discussed here. 

\subsubsection{Error evolution in the version of MAOOAM with a channel ocean} 

\begin{figure}
\centering
\includegraphics[width=1\textwidth]{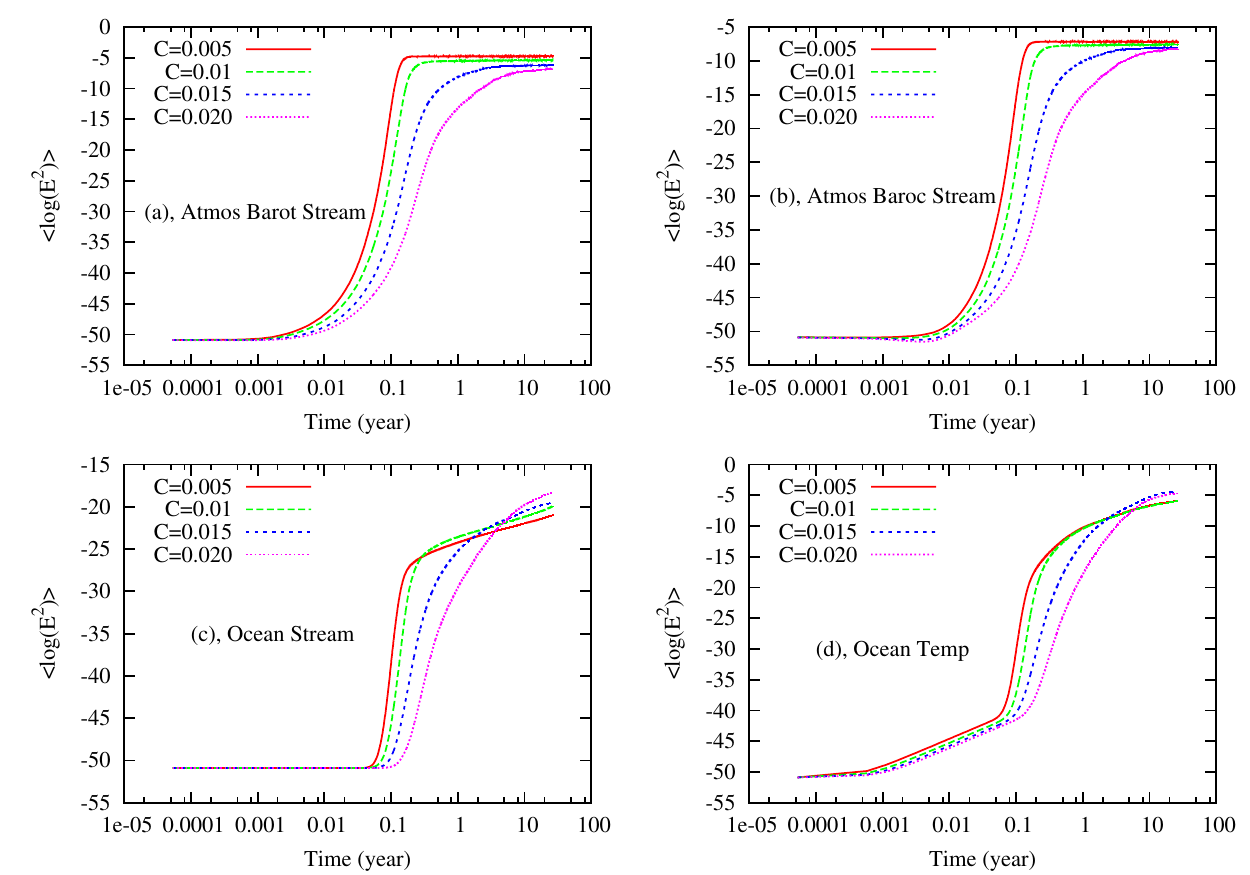}
\caption{Evolution of the averaged error measured using the logarithmic norm for (a) the atmospheric barotropic streamfunction; (b) the atmospheric baroclinic streamfunction; (c) the ocean
streamfunction; and (d) the ocean temperature. The different curves correspond to different values of the friction coefficient $C$. The parameter values used here are $h=1000$ m, 
$C_o=350$ W m$^{-2}$, $n=1.7$.}
\label{Figerr1}
\end{figure}

Equation \ref{expnorm0} is now computed for the version of the model with 40 variables presented in section \ref{sec:maosoamdescription}, with 1000 realizations starting from different initial states
on the attractor of the system. Figure \ref{Figerr1} displays the error evolution for the four fields of the model, namely 
the barotropic and baroclinic streamfunctions
of the atmosphere, the ocean temperature and the ocean streamfunction, for the parameters $h=1000$ m, $C_o=350$ W m$^{-2}$, $n=1.7$. The different curves in each panel
correspond to different values of the friction coefficient $C$. 

For a value of the friction coefficient, $C=0.005$ kg m$^{-2}$ s$^{-1}$, the error amplifies rapidly for the barotropic and baroclinic streamfunctions in
the atmosphere, with a saturation level reached before 0.2 years (Figure \ref{Figerr1}, panels (a) and (b)). For the ocean fields, the picture is different
with a very stable error until 0.1 year, after which it increases considerably concomitantly to the error amplification in the 
atmosphere. This phase is then followed by a slow increases for long lead times up to more than 20 years (Figure \ref{Figerr1}, panels (c) and (d)). 
This slow increase of the error in the ocean starts when the error in the atmosphere has already reached its saturation level.     
A similar picture has been found in the VDDG model version \cite{Vannitsem2016, Vannitsem2017b} for small friction coefficients, with a limited predictability to about a month in the atmosphere and a long term predictability in the ocean. 

When the friction coefficient is increased, the initial error amplification is slower due to the smaller value of the dominant
Lyapunov exponent (Figure \ref{Figlyap1}), explaining the slight shift of the error growth curve to the right. Besides this shift, the error behaviour
in the atmosphere and the ocean is similar as for $C=0.005$ kg m$^{-2}$ s$^{-1}$. The saturation of the error is however substantially modified
when $C$ is further increased. The saturation levels for the atmospheric fields are reached after about 2 years for $C$=0.015 kg m$^{-2}$ s$^{-1}$
and after about 20 years for $C$=0.02 kg m$^{-2}$ s$^{-1}$. In the ocean the error is still increasing after 25 years. This result reveals that the
friction coefficient $C$ plays an important role in the long term predictability of the atmospheric fields. 

\begin{figure}
\centering
\includegraphics[width=1\textwidth]{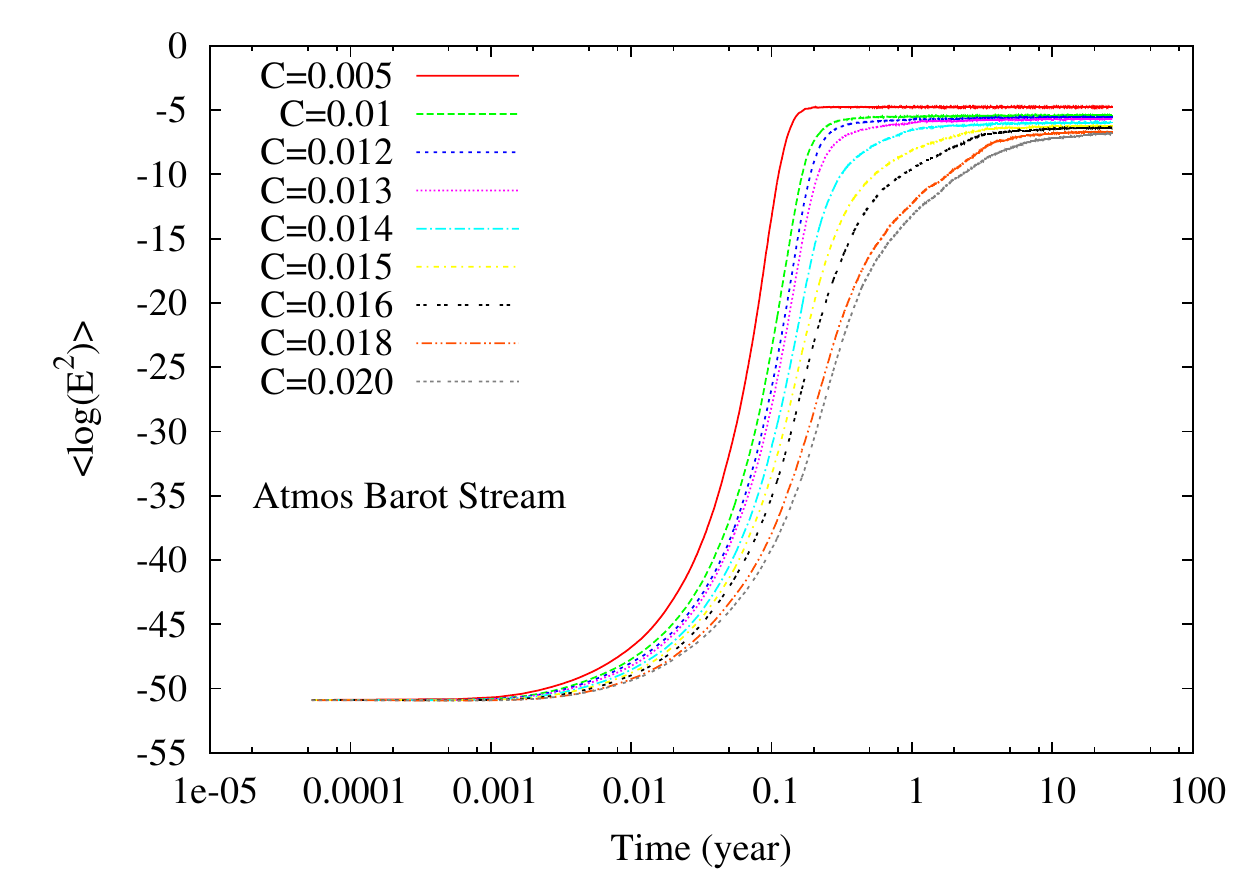}
\caption{Evolution of the averaged error measured using the logarithmic norm for the atmospheric barotropic streamfunction for different values of $C$. The parameter values used here are $h=1000$ m,
$C_o=350$ W m$^{-2}$, $n=1.7$.}
\label{Figerr2}
\end{figure}

Several other values of $C$ have been explored as illustrated in Figure \ref{Figerr2}. In this figure the error for the barotropic 
streamfunction fields saturates at larger and larger lead times as a function of $C$, for values above $C$=0.01 kg m$^{-2}$ s$^{-1}$. For $C$=0.02 kg m$^{-2}$ s$^{-1}$, the
error has not reach its saturation level after about 20 years yet. This feature is reminiscent of the change of long term predictability
experienced in the VDDG model \citep{Vannitsem2015a, Vannitsem2016, Vannitsem2017b} for large values of friction, with a long term 
predictability associated with the presence of an unstable periodic orbit around which the attractor is organized. The dynamics 
around this unstable periodic orbit however only appears after a chaos-to-chaos transition. 
This modification is visible in the abrupt change of the dominant Lyapunov exponents as illustrated in Figure 10 
of \cite{Vannitsem2017b}. One can wonder whether a similar change is experienced in the present model version. A first remark is
that there is no such drastic change of the dominant Lyapunov exponent in the present model version as shown by the green filled 
circles in panel (b) of Figure \ref{Figlyap1} for $n=1.7$, although the variations of the dominant Lyapunov exponent seem 
larger when $C$ is increased from 0.01 to 0.02 kg m$^{-2}$ s$^{-1}$.

In order to understand this property, it is necessary to go back to the dynamics described in Section \ref{solution}. 
For large values of the friction coefficient $C$, the solution of the model from time to time makes
an excursion in the vicinity of an unstable periodic orbit, for which the atmosphere is almost frozen and closely connected with the motion of the ocean. In
such a situation the solution is predictable for quite a long time. In the statistics of the error evolution, this leads to a slow
saturation of the error for long times. This intermittent behaviour lies at the origin of the long-term predictability of the atmosphere illustrated
in Figure (\ref{Figerr2}). For small values of $C$, there is no excursion close to such a periodic solution, leaving   
the atmosphere behaving with a high level of chaoticity. 

\begin{figure}
\centering
\includegraphics[width=1\textwidth]{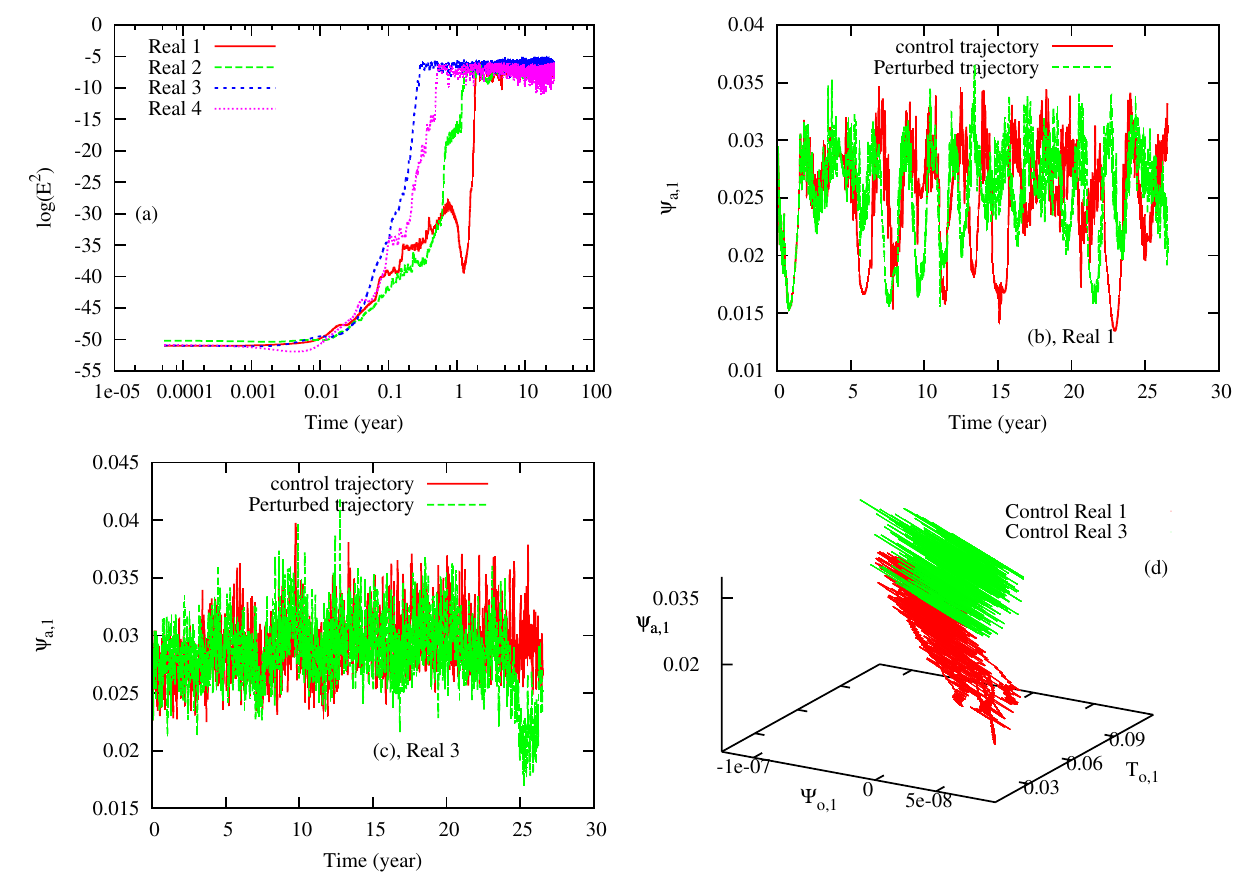}
\caption{(a) Four realisations of the error evolution for $C=0.016$ kg m$^{-2}$ s$^{-1}$, $h=1000$ m, $C_o=350$ W m$^{-2}$, $n=1.7$; (b) control and perturbed trajectories for realisation 1;
(c) control and perturbed trajectories for realisation 3; and (d) control trajectories of realisations 1 and 3 in a three-dimensional projection of the state space.}
\label{Figerr2-real}
\end{figure}

In order to further elucidate how excursions close to the periodic solution is inducing long-term predictability, the behaviour of the
solutions for several realizations of the error evolution have been investigated in more details. Figure \ref{Figerr2-real}a displays
several realizations of the error evolution. The first one shows a rapid increase of the error which saturates after about 0.2 years, but the others
display saturation only after 0.5 years, 1.5 years, and 2 years. This clearly suggests that in some circumstances long term 
predictability is present in the coupled system. If one looks at the two extreme cases, realizations 1 and 3, at panels (b) and (c) of Figure
\ref{Figerr2-real} where the control and perturbed dynamics of the variable $\psi_{a,1}$ are displayed, one can realize that a
very different dynamics is found. For Realization 1, an oscillating behaviour (but still chaotic) is present suggesting that the
dynamics is partly driven by the existence of an (unstable) periodic orbit. This evolution is accompanied with a small sensitivity to initial
states as reflected by the superposition of the control and the perturbed solutions for about up to 2 years. Note also that both 
solutions are still close to each other along the oscillating pattern up to 5 years, indicating that there is still a good potential
for prediction up to that lead time (this phase is less clear in the error evolution since we used the logarithm of the error). 
For realization 3, the picture is very different with much less apparent oscillations, and much less concordance between the two
trajectories. In the latter case, the long term predictability is very low. 

Finally it is interesting to see on which part of the attractor these two realizations are located. Figure \ref{Figerr2-real}d shows       
the 3-dimensional projection of the solutions on which the two realisations 1 and 3 of the error evolution are based. These are well
separated in state space, with realisation 3 spanning the lower right part of the projected attractor. This lower part has been identified
in the video S2 as the region where the periodic orbit should lie. 

So it clearly appears that the long term predictability of the solutions is associated with the intermittent excursions of the trajectories
in the vicinity of an unstable periodic orbit. As illustrated in video S2, these excursions occur apparently irregularly 
suggesting intermittent transitions in the vicinity of the unstable periodic orbit. This contrasts with the much more regular wandering behaviour around the
unstable periodic orbit found in the context of the VDDG model. 

\begin{figure}
\centering
\includegraphics[width=1\textwidth]{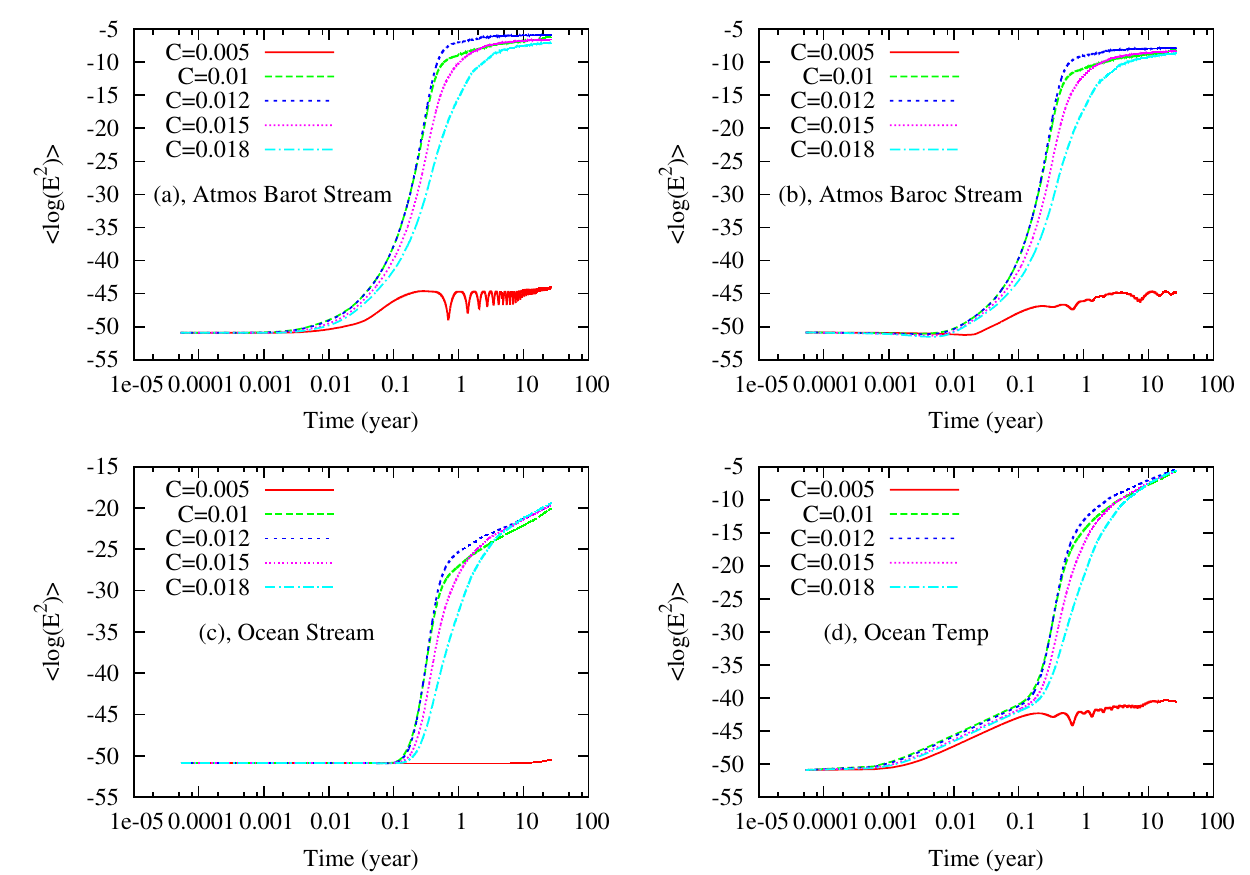}
\caption{Evolution of the averaged error measured using the logarithmic norm for (a) the atmospheric barotropic streamfunction; (b) the atmospheric baroclinic streamfunction; (c) the ocean
streamfunction; and (d) the ocean temperature. The different curves correspond to different values of the friction coefficient $C$. The parameter values used here are $h=1000$ m, 
$C_o=350$ W m$^{-2}$, $n=1.5$.}
\label{Figerr3}
\end{figure}

Let us now turn to the ocean-friction induced chaos as found with $n=1.5$. Figure \ref{Figerr3} displays the error evolution
for different values of $C$. For $C$ in the chaotic regime, a similar evolution as for $n=1.7$ is found, with a rapid initial
amplification and then a saturation phase. For most of the cases explored, the saturation level is only reach at very long lead times,
the only case with a short predictability period (still beyond a year) is the one for $C=0.012$ kg m$^{-2}$ s$^{-1}$. To interpret this long term 
predictability, let us figure out what kind of dynamics is taking place by looking at the different videos made with $n=1.5$ (videos S5-S7).
As for $n=1.7$, the solutions behave chaotically with intermittent excursions close to a periodic solution with a long period,
leading to a slow saturation of the error for long lead times.

\begin{figure}
\centering
\includegraphics[width=1\textwidth]{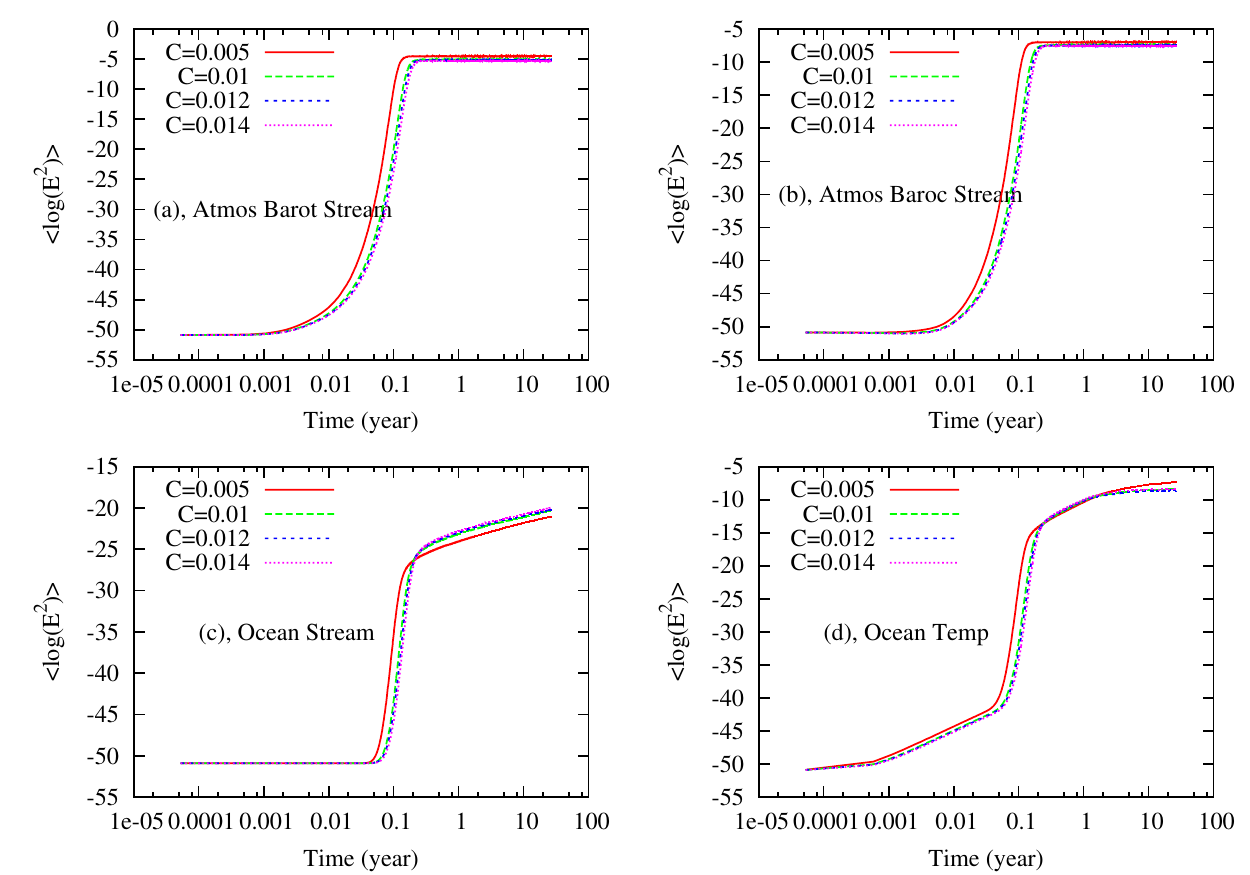}
\caption{Evolution of the averaged error measured using the logarithmic norm for (a) the atmospheric barotropic streamfunction; (b) the atmospheric baroclinic streamfunction; (c) the ocean
streamfunction; and (d) the ocean temperature. The different curves correspond to different values of the friction coefficient $C$. The parameter values used here are $h=100$ m, 
$C_o=350$ W m$^{-2}$, $n=1.7$.}
\label{Figerr4}
\end{figure}

\begin{figure}
\centering
\includegraphics[width=1\textwidth]{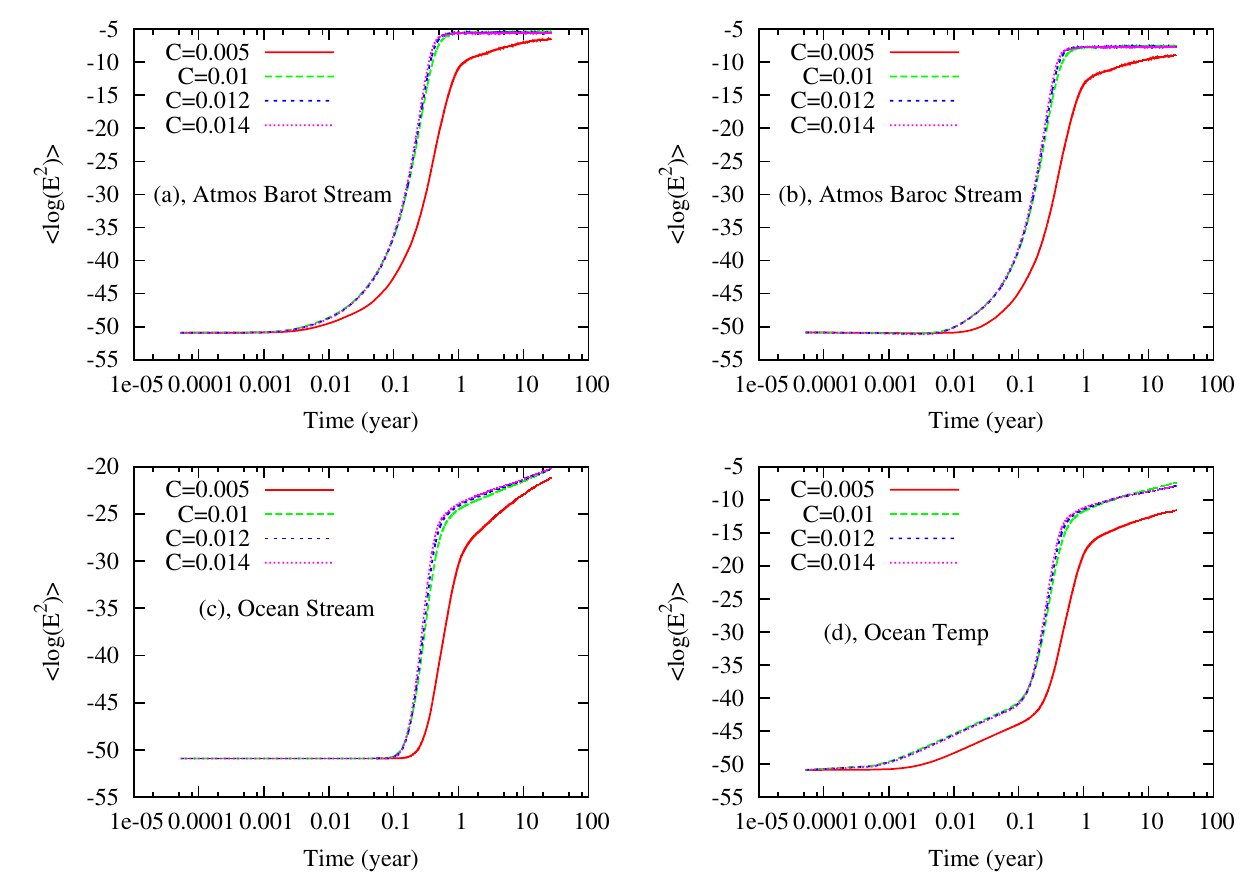}
\caption{Evolution of the averaged error measured using the logarithmic norm for (a) the atmospheric barotropic streamfunction; (b) the atmospheric baroclinic streamfunction; (c) the ocean
streamfunction; and (d) the ocean temperature. The different curves correspond to different values of the friction coefficient $C$. The parameter values used here are $h=100$ m, 
$C_o=350$ W m$^{-2}$, $n=1.5$.}
\label{Figerr5}
\end{figure}

A similar analysis of the error evolution is conducted for another depth of the ocean layer, $h=100$ m (Figures \ref{Figerr4}-\ref{Figerr5}).
In this situation, the evolution of the error in the atmosphere does not display any long term predictability, except for
$n=1.5$ and $C=0.005$ kg m$^{-2}$ s$^{-1}$. The intermittent behaviour found for $h=1000$ m does not emerge from the dynamics anymore. 
This contrasting behaviour suggests that the conditions for getting long term predictability are not always met and even
when a strong coupling (large friction and large heat transfer between the components of the system) exists, the predictability
is still limited to the typical time scale associated with the inverse of the dominant Lyapunov exponent.

In summary, the emergence of long term predictability in the present model version considerably differs 
from the behaviour found in the closed basin version of the model \citep{Vannitsem2015a, Vannitsem2015b, Vannitsem2017b}. 
In the latter work as already mentioned, a chaos-to-chaos bifurcation was identified leading to the development of a chaotic attractor around an unstable periodic orbit at the origin
of a low-frequency variability. The solutions are then wandering around the unstable periodic orbit in a similar way as in simpler systems \citep{Tel2006, Wernecke2017}. 
In the present system, an unstable periodic orbit is also involved, but the solution of the system is only experiencing
intermittent excursions in the vicinity of this orbit. These excursions also last longer when the value of $C$ is increased.   

\section{Conclusions}

A new geometry for the integration of the coupled ocean-atmosphere model, MAOOAM, has been implemented with a channel flow for both the atmosphere and the ocean
(periodic boundary conditions in the zonal direction for both model components). This new model version mimics the conditions that could be present around 
Antarctica. The predictability properties that may arise from the interaction between the ocean and the 
atmosphere are explored based on both the computation of the Lyapunov exponents and the long-term convergence of the mean square error toward a plateau, signature
of the loss of predictability.    

A first important result is that the interaction with the ocean can either induce chaos when the aspect ratio between the meridional and zonal length scales is small, or
suppress chaos when the aspect ratio is large. This feature has been found to be robust to modifications of the depth of the ocean and the radiative input into the
system. The ocean-induced chaotic regime may however be simply an artifact of the truncation of the spatial fields to their low-order versions, here limited to 10
Fourier modes. 


A second remarkable result of the analysis is that long-term predictability as measured by the mean square error evolution 
is not a robust feature emerging from the coupling, as it was the case in the coupled ocean-atmosphere model with closed boundaries for the ocean 
\cite{Vannitsem2015a, Vannitsem2015b, Vannitsem2017b}. To have long term predictability (in the mean), some specific parameter values (the depth of the ocean, the friction
coefficient) should be set in such a way that the solution of the model operates intermittent excursions in the vicinity of an unstable periodic orbit. 
This clearly demonstrates that emergence of low-frequency variability and the associated long-term predictability is not straightforwardly linked to the coupling
between the different components of the climate system. For the interaction with the ocean as discussed here, this emergence clearly depends on the basin geometry in reduced-order models like MAOOAM.  

Higher-resolution model versions should be explored in order to get confidence in these results. A preliminary analysis with model versions of higher dimensions with up to 312 variables, half for the atmosphere and half for the ocean, has been performed. Figure 10 illustrates the results for $h=1000$ m, $C_o$= 350 W m$^{-2}$, n=1.7, and different values of $C$ arbitrarily taken. When $C$ is small, a highly erratic behaviour is found, while for larger values of $C$ the behaviour displays strong similarities with the dynamics at lower resolutions shown in Figure 1, with an erratic decrease of the dominant ocean temperature mode close to 0 (see panel (c)). This
suggests that the dynamics found for the reduced-order model is a generic dynamics that can be found at higher resolutions. The
mechanism behind this dynamics is therefore worth investigating in more realistic models like the one developed by \citet{Hogg2003} and also in realistic climate models.

\begin{figure}
\centering
\includegraphics[width=1\textwidth]{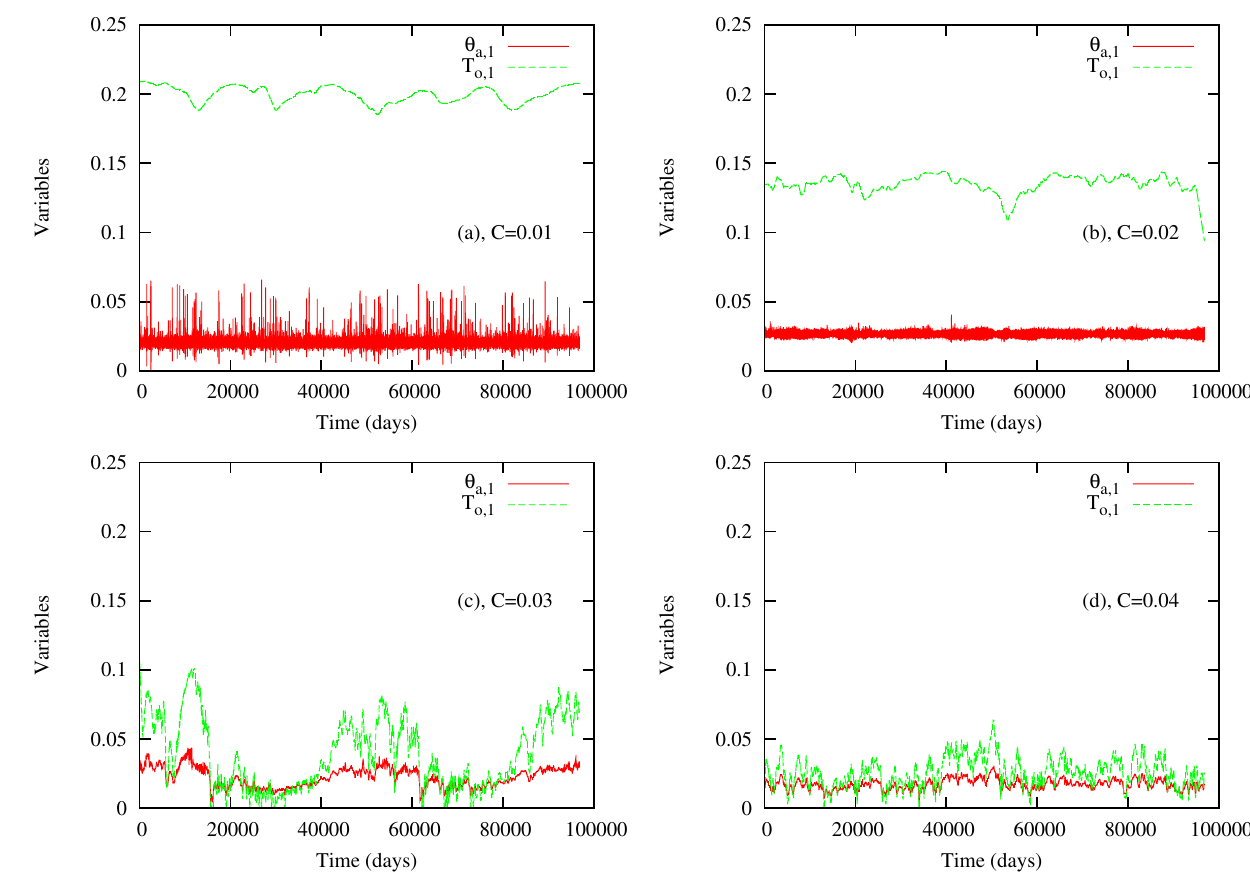}
\caption{Time series of $\theta_{a,1}$ and $T_{o,1}$ with a higher resolution model version with 312 variables for (a) $C=0.01$ kg m$^{-2}$ s$^{-1}$; (b) $C=0.02$ kg m$^{-2}$ s$^{-1}$; (c) $C=0.03$ kg m$^{-2}$ s$^{-1}$ and (d) $C=0.04$ kg m$^{-2}$ s$^{-1}$. The other parameters are $n=1.7$, $C_o=350$ W m$^{-2}$ and $h=1000$ m. Nondimensional units are used for the variables.}
\label{Fighires}
\end{figure}

The present open-channel ocean model has a horizontal geometry with some similarities with the open-channel structure of the Southern Ocean. However the homogeneous vertical and horizontal structure of the
model is far from being appropriate to describe the Southern ocean since strong horizontal variations of the density are present inducing the development of the 
Antarctic Circumpolar Current (ACC), \citep{Vallis2006}. The model is thus not able to describe the ACC appropriately. One can therefore wonder whether 
our analysis is relevant for the actual Southern Ocean. This can be addressed by modifying the ocean dynamics allowing for horizontal gradient of densities and the development of 
convective instabilities. This analysis is planned in the future 
 along the lines of the ocean modelling approach of \citet{Hogg2003} for instance.  

Finally, in the VDDG and the current model versions, the long-term predictability is associated with the presence of an unstable periodic orbit whose origin can be traced back to
the interaction between the ocean and the atmosphere. This long-term predictability mechanism was already discovered in much simpler "single scale" systems 
\cite{Tel2006, Gros2015, Wernecke2017} and the dynamics developing in this context is referred to as {\it partially predictable chaos}. Our results are 
reminiscent of this mechanism but here in a multiscale system with very long periods of motion along the unstable periodic orbits. The current results, however, point toward
the possibility of having different routes to long-term predictability. A first one has been obtained in \cite{Vannitsem2015a, Vannitsem2015b, Vannitsem2017b} for the closed ocean basin
model with a wandering along the unstable periodic orbit, and second as in the present model configuration, an intermittent transition from a highly chaotic attracting set to the vicinity of an unstable periodic orbit.
In the latter case, only some forecasts inherit of the long-term predictability properties and the difficulty is to know when this transition will occur. Finding
precursors of this transition is essential in order to assess the potential predictability of such forecasts, like for instance when searching for precursors of atmospheric
blocking \citep{Vautard1990, Matsueda2018} or the precursors of specific ENSO events \citep{Duan2004}.    

\section{Supporting information}

Seven videos were prepared to visualize the solutions of the model for different values of the parameters, and are provided as supplementary material. They are listed from video S1 to video S7. Table \ref{tab:vid} lists the parameter values and the digital object identifiers (DOIs) for each video. The other parameters are fixed at $C_o=350$ W m$^{-2}$ and $h=1000$ m for these simulations.

\begin{table}
\caption{List of videos\label{tab:vid}}
\begin{tabular}{lllr}
\hline
    Video & n & C (kg m$^{-2}$ s$^{-1}$) & DOI \\
    S1  & 1.7 & 0.010 &\url{https://doi.org/10.5446/39179} \\
    S2  & 1.7 & 0.016 &\url{https://doi.org/10.5446/39180} \\
    S3  & 1.7 & 0.020 &\url{https://doi.org/10.5446/39181} \\
    S4  & 1.7 & 0.027 &\url{https://doi.org/10.5446/39182} \\
    S5  & 1.5 & 0.010 &\url{https://doi.org/10.5446/39183} \\
    S6  & 1.5 & 0.012 &\url{https://doi.org/10.5446/39184} \\
    S7  & 1.5 & 0.016 &\url{https://doi.org/10.5446/39185} \\
\hline
  \end{tabular}
\end{table}

\bibliographystyle{abbrvnat}
\bibliography{maosoam}

\begin{thebibliography}{41}
\providecommand{\natexlab}[1]{#1}
\providecommand{\url}[1]{\texttt{#1}}
\expandafter\ifx\csname urlstyle\endcsname\relax
  \providecommand{\doi}[1]{doi: #1}\else
  \providecommand{\doi}{doi: \begingroup \urlstyle{rm}\Url}\fi

\bibitem[Boer(1994)]{Boer1994}
G.~J. Boer.
\newblock Predictability regimes in atmospheric flow.
\newblock \emph{Monthly Weather Review}, 122\penalty0 (10):\penalty0
  2285--2295, 1994.
\newblock \doi{10.1175/1520-0493(1994)122<2285:PRIAF>2.0.CO;2}.

\bibitem[Brachet et~al.(2012)Brachet, Codron, Feliks, Ghil, Le~Treut, and
  Simonnet]{Brachet2012}
S.~Brachet, F.~Codron, Y.~Feliks, M.~Ghil, H.~Le~Treut, and E.~Simonnet.
\newblock {Atmospheric circulations induced by a midlatitude SST front: A GCM
  study}.
\newblock \emph{Journal of Climate}, 25\penalty0 (6):\penalty0 1847--1853,
  2012.
\newblock \doi{10.1175/JCLI-D-11-00329.1}.

\bibitem[Cehelsky and Tung(1987)]{Cehelsky1987}
P.~Cehelsky and K.~K. Tung.
\newblock {Theories of multiple equilibria and weather regimes -- A critical
  reexamination. Part II: Baroclinic two-layer models}.
\newblock \emph{Journal of the Atmospheric Sciences}, 44\penalty0
  (21):\penalty0 3282--3303, 1987.

\bibitem[Charney and Straus(1980)]{Charney1980}
J.~G. Charney and D.~M. Straus.
\newblock Form-drag instability, multiple equilibria and propagating planetary
  waves in baroclinic, orographically forced, planetary wave systems.
\newblock \emph{Journal of the Atmospheric Sciences}, 37\penalty0 (6):\penalty0
  1157--1176, 1980.

\bibitem[Chen(1989)]{Chen1989}
W.~Y. Chen.
\newblock {Estimate of dynamical predictability from NMC DERF experiments}.
\newblock \emph{Monthly Weather Review}, 117\penalty0 (6):\penalty0 1227--1236,
  1989.

\bibitem[Czaja and Frankignoul(2002)]{Czaja2002}
A.~Czaja and C.~Frankignoul.
\newblock {Observed impact of Atlantic SST anomalies on the North Atlantic
  Oscillation}.
\newblock \emph{Journal of Climate}, 15\penalty0 (6):\penalty0 606--623, 2002.

\bibitem[Dalcher and Kalnay(1987)]{Dalcher1987}
A.~Dalcher and E.~Kalnay.
\newblock {Error growth and predictability in operational ECMWF forecasts}.
\newblock \emph{Tellus A: Dynamic Meteorology and Oceanography}, 39\penalty0
  (5):\penalty0 474--491, 1987.

\bibitem[{De Cruz} et~al.(2016){De Cruz}, Demaeyer, and Vannitsem]{DeCruz2016}
L.~{De Cruz}, J.~Demaeyer, and S.~Vannitsem.
\newblock {The Modular Arbitrary-Order Ocean-Atmosphere Model: MAOOAM v1.0}.
\newblock \emph{Geoscientific Model Development}, 9\penalty0 (8):\penalty0
  2793--2808, 2016.

\bibitem[{De Cruz} et~al.(2018){De Cruz}, Schubert, Demaeyer, Lucarini, and
  Vannitsem]{DeCruz2018}
L.~{De Cruz}, S.~Schubert, J.~Demaeyer, V.~Lucarini, and S.~Vannitsem.
\newblock {Exploring the Lyapunov instability properties of high-dimensional
  atmospheric and climate models}.
\newblock \emph{Nonlinear Processes in Geophysics}, 25\penalty0 (2):\penalty0
  387--412, 2018.
\newblock \doi{10.5194/npg-25-387-2018}.

\bibitem[Demaeyer and Vannitsem(2018)]{Demaeyer2018}
J.~Demaeyer and S.~Vannitsem.
\newblock Comparison of stochastic parameterizations in the framework of a
  coupled ocean-atmosphere model.
\newblock \emph{Nonlinear Processes in Geophysics}, 25\penalty0 (3):\penalty0
  605--631, 2018.
\newblock \doi{10.5194/npg-25-605-2018}.

\bibitem[Duan et~al.(2004)Duan, Mu, and Wang]{Duan2004}
W.~Duan, M.~Mu, and B.~Wang.
\newblock {Conditional nonlinear optimal perturbations as the optimal
  precursors for El Niño--Southern Oscillation events}.
\newblock \emph{Journal of Geophysical Research: Atmospheres}, 109\penalty0
  (D23), 2004.
\newblock \doi{10.1029/2004JD004756}.

\bibitem[Eckmann and Ruelle(1985)]{Eckmann1985}
J.~Eckmann and D.~Ruelle.
\newblock {Ergodic theory of chaos and strange attractors}.
\newblock \emph{Reviews of Modern Physics}, 57\penalty0 (3):\penalty0 617--656,
  7 1985.
\newblock ISSN 0034-6861.
\newblock \doi{10.1103/RevModPhys.57.617}.

\bibitem[Feliks et~al.(2007)Feliks, Ghil, and Simonnet]{Feliks2007}
Y.~Feliks, M.~Ghil, and E.~Simonnet.
\newblock Low-frequency variability in the midlatitude baroclinic atmosphere
  induced by an oceanic thermal front.
\newblock \emph{Journal of the Atmospheric Sciences}, 64\penalty0 (1):\penalty0
  97--116, 2007.

\bibitem[Gros(2015)]{Gros2015}
C.~Gros.
\newblock \emph{Complex and adaptive dynamical systems: A primer}.
\newblock Springer, Switzerland, 2015.

\bibitem[Hogg et~al.(2003)Hogg, Dewar, Killworth, and Blundell]{Hogg2003}
A.~M.~C. Hogg, W.~K. Dewar, P.~D. Killworth, and J.~R. Blundell.
\newblock A quasi-geostrophic coupled model (q-gcm).
\newblock \emph{Monthly Weather Review}, 131\penalty0 (10):\penalty0
  2261--2278, 2003.
\newblock \doi{10.1175/1520-0493(2003)131<2261:AQCMQ>2.0.CO;2}.

\bibitem[Houghton(1986)]{Houghton1986}
J.~T. Houghton.
\newblock \emph{The Physics of Atmospheres}.
\newblock Cambridge University Press, Cambridge, 1986.

\bibitem[Kalnay(2003)]{Kalnay2003}
E.~Kalnay.
\newblock \emph{Atmospheric Modeling, Data Assimilation, and Predictability}.
\newblock Cambridge University Press, Cambridge, UK, 2003.

\bibitem[Kravtsov et~al.(2007)Kravtsov, Dewar, Berloff, McWilliams, and
  Ghil]{Kravtsov2007}
S.~Kravtsov, W.~K. Dewar, P.~Berloff, J.~C. McWilliams, and M.~Ghil.
\newblock A highly nonlinear coupled mode of decadal variability in a
  mid-latitude ocean--atmosphere model.
\newblock \emph{Dynamics of Atmospheres and Oceans}, 43\penalty0 (3):\penalty0
  123--150, 2007.

\bibitem[Kuptsov and Parlitz(2012)]{Kuptsov2012}
P.~V. Kuptsov and U.~Parlitz.
\newblock {Theory and Computation of Covariant Lyapunov Vectors}.
\newblock \emph{Journal of Nonlinear Science}, 22\penalty0 (5):\penalty0
  727--762, 3 2012.
\newblock ISSN 0938-8974.
\newblock \doi{10.1007/s00332-012-9126-5}.

\bibitem[Lorenz(1982)]{Lorenz1982}
E.~N. Lorenz.
\newblock Atmospheric predictability experiments with a large numerical model.
\newblock \emph{Tellus}, 34\penalty0 (6):\penalty0 505--513, 1982.

\bibitem[L’H{\'e}v{\'e}der et~al.(2015)L’H{\'e}v{\'e}der, Codron, and
  Ghil]{Lheveder2015}
B.~L’H{\'e}v{\'e}der, F.~Codron, and M.~Ghil.
\newblock Impact of anomalous northward oceanic heat transport on global
  climate in a slab ocean setting.
\newblock \emph{Journal of Climate}, 28\penalty0 (7):\penalty0 2650--2664,
  2015.

\bibitem[Marshall et~al.(2001)Marshall, Johnson, and Goodman]{Marshall2001}
J.~Marshall, H.~Johnson, and J.~Goodman.
\newblock {A study of the interaction of the North Atlantic Oscillation with
  ocean circulation}.
\newblock \emph{Journal of Climate}, 14\penalty0 (7):\penalty0 1399--1421,
  2001.

\bibitem[Matsueda and Palmer(2018)]{Matsueda2018}
M.~Matsueda and T.~N. Palmer.
\newblock {Estimates of flow-dependent predictability of wintertime
  Euro-Atlantic weather regimes in medium-range forecasts}.
\newblock \emph{Quarterly Journal of the Royal Meteorological Society},
  144\penalty0 (713):\penalty0 1012--1027, 2018.

\bibitem[Minobe et~al.(2008)Minobe, Kuwano-Yoshida, Komori, Xie, and
  Small]{Minobe2008}
S.~Minobe, A.~Kuwano-Yoshida, N.~Komori, S.-P. Xie, and R.~J. Small.
\newblock {Influence of the Gulf Stream on the troposphere}.
\newblock \emph{Nature}, 452\penalty0 (7184):\penalty0 206--209, 2008.

\bibitem[Nicolis et~al.(1995)Nicolis, Vannitsem, and Royer]{Nicolis1995}
C.~Nicolis, S.~Vannitsem, and J.-F. Royer.
\newblock Short-range predictability of the atmosphere: Mechanisms for
  superexponential error growth.
\newblock \emph{Quarterly Journal of the Royal Meteorological Society},
  121\penalty0 (523):\penalty0 705--722, 1995.

\bibitem[Parker and Chua(1989)]{Parker1989}
T.~S. Parker and L.~Chua.
\newblock \emph{Practical numerical algorithms for chaotic systems}.
\newblock Springer Verlag, New York, 1989.

\bibitem[Philander(1990)]{Philander1990}
S.~G.~H. Philander.
\newblock \emph{El Niño, La Niña, and the Southern Oscillation}, volume~46 of
  \emph{International Geophysics}.
\newblock Academic Press, San Diego, 1990.

\bibitem[Robertson et~al.(2000)Robertson, Ghil, and Latif]{Robertson2000}
A.~W. Robertson, M.~Ghil, and M.~Latif.
\newblock Interdecadal changes in atmospheric low-frequency variability with
  and without boundary forcing.
\newblock \emph{Journal of the Atmospheric Sciences}, 57\penalty0 (8):\penalty0
  1132--1140, 2000.

\bibitem[Savijarvi(1995)]{Savijarvi1995}
H.~Savijarvi.
\newblock Error growth in a large numerical forecast system.
\newblock \emph{Monthly Weather Review}, 123\penalty0 (1):\penalty0 212--221,
  1995.

\bibitem[Stuecker et~al.(2013)Stuecker, Timmermann, Jin, McGregor, and
  Ren]{Stuecker2013}
M.~F. Stuecker, A.~Timmermann, F.-F. Jin, S.~McGregor, and H.-L. Ren.
\newblock {A combination mode of the annual cycle and the El Ni{\~n}o/Southern
  Oscillation}.
\newblock \emph{Nature Geoscience}, 6\penalty0 (7):\penalty0 540, 2013.

\bibitem[T{\'e}l and Gruiz(2006)]{Tel2006}
T.~T{\'e}l and M.~Gruiz.
\newblock \emph{Chaotic dynamics: an introduction based on classical
  mechanics}.
\newblock Cambridge University Press, Cambridge, UK, 2006.

\bibitem[Vallis(2006)]{Vallis2006}
G.~K. Vallis.
\newblock \emph{Atmospheric and oceanic fluid dynamics: fundamentals and
  large-scale circulation}.
\newblock Cambridge University Press, Cambridge, UK, 2006.

\bibitem[Van~der Avoird et~al.(2002)Van~der Avoird, Dijkstra, Nauw, and
  Schuurmans]{VanderAvoird2002}
E.~Van~der Avoird, H.~Dijkstra, J.~Nauw, and C.~Schuurmans.
\newblock Nonlinearly induced low-frequency variability in a midlatitude
  coupled ocean-atmosphere model of intermediate complexity.
\newblock \emph{Climate Dynamics}, 19\penalty0 (3):\penalty0 303--320, 2002.

\bibitem[Vannitsem(2015)]{Vannitsem2015b}
S.~Vannitsem.
\newblock {The role of the ocean mixed layer on the development of the North
  Atlantic Oscillation: A dynamical system's perspective}.
\newblock \emph{Geophysical Research Letters}, 42\penalty0 (20):\penalty0
  8615--8623, 2015.

\bibitem[Vannitsem(2017)]{Vannitsem2017b}
S.~Vannitsem.
\newblock Predictability of large-scale atmospheric motions: {Lyapunov}
  exponents and error dynamics.
\newblock \emph{Chaos: An Interdisciplinary Journal of Nonlinear Science},
  27\penalty0 (3):\penalty0 32101, 3 2017.
\newblock ISSN 1054-1500.
\newblock \doi{10.1063/1.4979042}.

\bibitem[Vannitsem and Ghil(2017)]{Vannitsem2017a}
S.~Vannitsem and M.~Ghil.
\newblock {Evidence of coupling in ocean-atmosphere dynamics over the North
  Atlantic}.
\newblock \emph{Geophysical Research Letters}, 44\penalty0 (4):\penalty0
  2016--2026, 2017.

\bibitem[Vannitsem and Lucarini(2016)]{Vannitsem2016}
S.~Vannitsem and V.~Lucarini.
\newblock Statistical and dynamical properties of covariant {Lyapunov} vectors
  in a coupled atmosphere-ocean model -- multiscale effects, geometric
  degeneracy, and error dynamics.
\newblock \emph{Journal of Physics A: Mathematical and Theoretical},
  49\penalty0 (22):\penalty0 224001, 2016.

\bibitem[Vannitsem and Nicolis(1994)]{Vannitsem1994}
S.~Vannitsem and C.~Nicolis.
\newblock {Predictability experiments on a simplified thermal convection model:
  The role of spatial scales}.
\newblock \emph{Journal of Geophysical Research}, 99\penalty0 (D5):\penalty0
  10377, 1994.
\newblock ISSN 0148-0227.
\newblock \doi{10.1029/94JD00248}.

\bibitem[Vannitsem et~al.(2015)Vannitsem, Demaeyer, {De Cruz}, and
  Ghil]{Vannitsem2015a}
S.~Vannitsem, J.~Demaeyer, L.~{De Cruz}, and M.~Ghil.
\newblock {Low-frequency variability and heat transport in a low-order
  nonlinear coupled ocean-atmosphere model}.
\newblock \emph{Physica D: Nonlinear Phenomena}, 309:\penalty0 71--85, 2015.
\newblock ISSN 01672789.
\newblock \doi{10.1016/j.physd.2015.07.006}.

\bibitem[Vautard(1990)]{Vautard1990}
R.~Vautard.
\newblock {Multiple weather regimes over the North Atlantic: Analysis of
  precursors and successors}.
\newblock \emph{Monthly Weather Review}, 118\penalty0 (10):\penalty0
  2056--2081, 1990.

\bibitem[Wernecke et~al.(2017)Wernecke, S{\'a}ndor, and Gros]{Wernecke2017}
H.~Wernecke, B.~S{\'a}ndor, and C.~Gros.
\newblock How to test for partially predictable chaos.
\newblock \emph{Scientific Reports}, 7\penalty0 (1):\penalty0 1087, 2017.

\end{thebibliography}

\end{document}